\documentclass[letter]{aa} 

\usepackage{graphicx}
\usepackage{txfonts}
\usepackage{color}

\begin{document}

   \title{Close-in giant-planet formation via in-situ gas accretion and their natal disk properties}

   \authorrunning{Y. Hasegawa, T. Y. M. Yu, and B. M. S. Hansen}
   \titlerunning{Close-in Giant Planets and In-situ Gas Accretion}

   \author{Yasuhiro Hasegawa
          \inst{1},
          Tze Yeung Mathew Yu,
          \inst{2}
          \and
          Bradley M. S. Hansen
          \inst{2}
          }

   \institute{Jet Propulsion Laboratory, California Institute of Technology, Pasadena, CA 91109, USA\\
              \email{yasuhiro.hasegawa@jpl.nasa.gov}
         \and
             Department of Physics \& Astronomy, University of California Los Angeles, Los Angeles, CA 90095, USA
             }

   \date{Received; accepted }

 
  \abstract
   {}
   {The origin of close-in Jovian planets is still elusive. 
   We examine the in-situ gas accretion scenario as a formation mechanism of these planets.}
   {We reconstruct natal disk properties from the occurrence rate distribution of close-in giant planets,
   under the assumption that the occurrence rate may reflect the gas accretion efficiency onto cores of these planets.}
   {We find that the resulting gas surface density profile becomes an increasing function of the distance from the central star with some structure at $r \simeq 0.1$ au.
   This profile is quite different from the standard minimum-mass solar nebula model,
   while our profile leads to better reproduction of the population of observed close-in super-Earths based on previous studies.
   We compute the resulting magnetic field profiles and 
   find that our profiles can be fitted by stellar dipole fields ($\propto r^{-3}$) in the vicinity of the central star and large-scale fields ($\propto r^{-2}$) at the inner disk regions,
   either if the isothermal assumption breaks down or if nonideal MHD effects become important.
   For both cases, the transition between these two profiles occurs at $r \simeq 0.1$ au, which corresponds to the period valley of giant exoplanets.}
   {Our work provides an opportunity to test the in-situ gas accretion scenario against disk quantities,
   which may constrain the gas distribution of the minimum-mass {\it extra}solar nebula.}


   \maketitle

\section{Introduction} \label{sec:intro}

\begin{figure}
\begin{center}
\includegraphics[width=8cm]{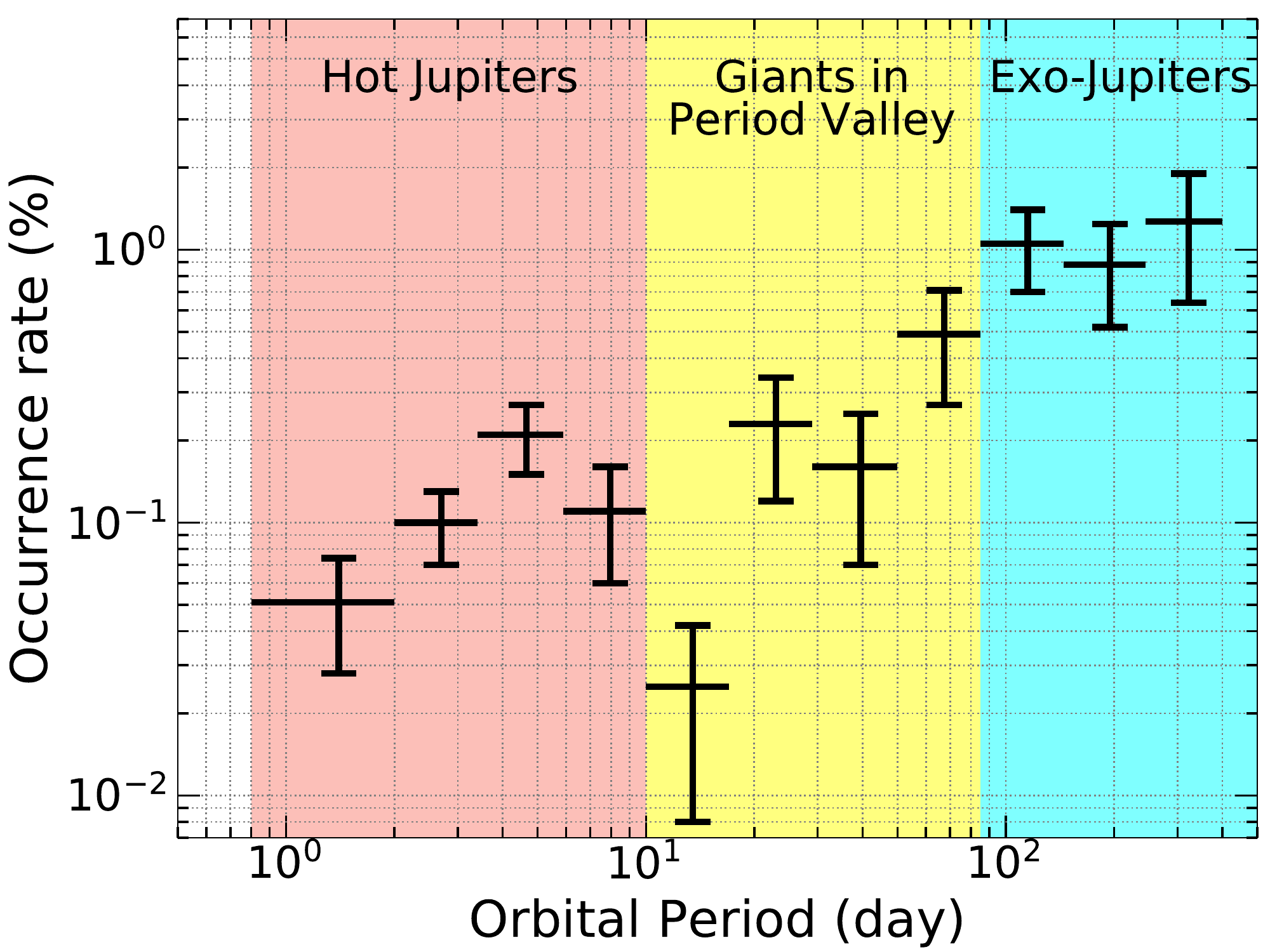}
\caption{Occurrence rates of close-in giant planets as a function of orbital period.
The horizontal bars represent the bin sizes used in estimating the occurrence rates,
while the vertical ones denote the error bars.
The data are taken from \citet{smt16}.}
\label{fig1}
\end{center}
\end{figure}

The dawn of exoplanetary science has come with the remarkable discovery of the Jovian-mass planet orbiting in the vicinity of its host, solar-type star \citep{mq95}.
The extrasolar planets referred to as hot Jupiters have stimulated a number of follow-up studies to understand their origins \citep[for a recent review]{dj18}.
The rapid accumulation of exoplanet populations has revealed that 
close-in Jovian planets including hot Jupiters are statistically rare \citep[also see Figure \ref{fig1}]{wf15}, and their origin is still unclear.

Three modes of forming close-in Jupiters are currently offered.
The first two modes are extensions of the canonical model: 
(proto)Jovian planets form beyond the snow line via the core accretion scenario \citep{p96}.
The new ingredients are radial movement of planets during or after the process of forming.
The first mode considers that  inward planetary migration caused by planet--disk interaction plays the dominant role in reproducing the current orbital distribution \citep{lbr96,kn12}.
The second one attributes such short orbital periods to planet--planet interaction coupled with stellar tides \citep{rf96,mpr10}.
The third mode is called in-situ gas accretion because it does not require large-scale radial movement \citep{bhl00,bbl16,bgg16}.
Identification of which mode(s) would be most responsible for shaping the occurrence rate of close-in Jupiters has not yet been achieved.

The lack of migration in the in-situ scenario implies that 
the formation efficiency of giant planets must be determined by the local properties of their natal disks.
Given that the occurrence rate of observed exoplanets should directly correlate with the formation efficiency,
it may be possible to invert exoplanet observations to infer the properties of the planet-forming disks.
Here, we attempt to conduct such a study, using the observational results of \citet{smt16}.
These authors employed {\it Kepler} transit observations combined with radial velocity follow-up
and obtained the occurrence rate distribution of close-in Jovian planets.
Figure \ref{fig1} shows that the distribution possesses some intriguing structure.
A similar structure was seen by \citet{pmw18} where only the {\it Kepler} data were used.\footnote{
Such a distribution is derived from planets that have radii of $\sim 4-24 R_{\oplus}$ (equivalently $\sim 3M_{\oplus}-80 M_{\rm Jupiter}$),
while planets residing in the lower and upper mass ends are rare.}
The fundamental hypothesis of this work is that 
the planet formation efficiency and hence the occurrence rate distribution are determined by the ability of giant planet cores to accrete the local disk gas.
Under this hypothesis,
we show below that the occurrence rate distribution can be converted to disk properties such as the gas surface density profile and magnetic field profile threading disks.
Thus, our work provides not only disk quantities for examining the in-situ gas accretion scenario,
but also a test bed for the gas distribution of the minimum mass {\it extra}solar nebula model \citep[cf.,][]{cl13}.

\section{Occurrence rate as a probe of inner disk structures} \label{sec:mod}

\subsection{Basic hypothesis} \label{sec:mod_1}

We begin with the basic hypothesis that can be drawn from the assumption that
the occurrence rate distribution ($f_{\rm OR} (r)$) of close-in giant exoplanets may be viewed as a proxy for the probability of giant planet formation at $r$.
Under the in-situ gas-accretion scenario,
this probability should scale with the efficiency for planetary cores to accrete the local disk gas.
Consequently, one can obtain the relationship between $f_{\rm OR}$ and inner disk properties (see below).

Gas accretion onto planets ($\dot{M}_{p}$) can be divided into three stages in the core accretion scenario \citep[e.g.,][]{mak12,hhv19}.
The first stage is envelope contraction around planetary cores characterized by the Kelvin-Helmholtz timescale.
Given that the disk gas contained within the Hill radius of the cores with $\ga 10 M_{\oplus}$ is a few Neptune masses at most at small orbital periods,
this stage would end quickly and is not important for $f_{\rm OR}$.
In the second stage, proto-Jovian planets keep accreting the disk gas, 
but the resulting accretion rate becomes too high and should eventually be limited by the {\it local} mass supply via disk evolution.
This stage is the so-called disk-limited gas accretion.
Such efficient gas accretion is needed for the formation of giant planets,
some of which would constitute the samples analyzed in \citet{smt16}.
When planets are massive enough to open up gaps in their gas disks,
$\dot{M}_{p}$ is determined by the gas coming from the polar direction \citep{msc14}.
While the final mass of planets should be determined during this stage,
the value does not represent the local disk properties.
Instead, the mass supply from outer disks over the gas disk lifetime of $\sim 10^6$ yr controls this quantity \citep{bhl00}.\footnote{
Under the steady state disk accretion model, 
$\dot{M}_{p}$ in this stage becomes an increasing function of $r$ \citep[e.g., see equations (8) and (9) in][]{hhv19},
which may be related to a radial distribution of the final mass of planets.}
Thus, the disk-limited gas accretion is the most likely determinant of the gas giant formation efficiency and hence $f_{\rm OR}$ under the in-situ scenario.
Mathematically, we formulate the disk-limited gas accretion rate as \citep{ti07}
\begin{eqnarray}
\label{eq:Mpdot_hydro}
\dot{M}_{p} & =          &  0.29 \left( \frac{H_{\rm g}}{r_p} \right)^{-2}  \left( \frac{M_p}{M_*} \right)^{4/3} \Sigma_{\rm g} r_p^2 \Omega  \\ \nonumber
                                   & \simeq &  1.5 \times 10^{-3}  \left( \frac{\alpha}{10^{-2}} \right)^{-1} \left( \frac{H_{\rm g}/r_p}{0.05} \right)^{-4}  \\ \nonumber
                                   & \times  &   \left( \frac{M_p}{10 M_{\oplus}} \right)^{4/3} \left( \frac{\dot{M}_{\rm d}}{10^{-8} M_{\odot} \mbox{ yr}^{-1}}  \right)       \frac{M_{\oplus}}{\mbox{yr}}.
\end{eqnarray}
This formula is derived from hydrodynamical simulations \citep{tw02} and reliably reproduces the simulation results that were run independently by different groups 
\citep[references therein]{hbi18}.
We note that the steady state disk accretion model is adopted in equation (\ref{eq:Mpdot_hydro}),
where the disk accretion rate ($\dot{M}_{\rm d}$) is given as  \citep{fkr02}
\begin{equation}
\label{eq:mdot}
\dot{M}_{\rm d} = 3 \pi \nu \Sigma_{\rm g},
\end{equation}
where $\Sigma_{\rm g}$ is the gas surface density, $\nu=\alpha c_{\rm s} H_{\rm g}$ is the effective viscosity, 
$c_{\rm s}(\propto T_{\rm d}^{1/2})$ is the sound speed, $T_{\rm d}$ is the disk temperature, $H_{\rm g}=c_{\rm s}/ \Omega$ is the pressure scale height, 
and $\Omega$ is the Keplerian angular velocity around a solar mass star.
The $\alpha$-parameterization is used for quantifying the efficiency of angular momentum transport in disks \citep{ss73}.

Consequently, our hypothesis can be written as
\begin{equation}
\label{eq:f_OR}
f_{\rm OR} (r) \equiv  k_{\rm OR} \frac{\dot{M}_{p}}{\dot{M}_{\rm d}} 
                        \simeq   0.46  k_{\rm OR} \left( \frac{\alpha}{10^{-2}} \right)^{-1} \left( \frac{H_{\rm g}/r_p}{0.05} \right)^{-4} \left( \frac{M_p}{10 M_{\oplus}} \right)^{4/3} \%, 
\end{equation}
where $k_{\rm OR}$ is an unknown factor related to the giant planet formation efficiency.
This equation shows that a higher $\dot{M}_p$ leads to a higher $f_{\rm OR}$.

Thus, one can use $f_{\rm OR}$ for specifying disk properties (either $\alpha$ or $T_{\rm d}$) under our hypothesis.

\subsection{Disk model} \label{sec:mod_2}

Determination of $T_{\rm d}$ becomes fundamental for exploring natal disk structures from $f_{\rm OR}$ (see equation (\ref{eq:f_OR})).
Here we describe how $T_{\rm d}$ is computed self-consistently.

Protoplanetary disks can be heated by two sources \citep{a11}: viscous heating and stellar irradiation.
Viscous heating originates from turbulent accretion stress
and generally becomes important in the inner region of the disk \citep{dccl98}.
On the other hand, stellar irradiation regulates the disk temperature in the outer region of disk \citep{cg97}.
This work focuses on the inner region of disk and hence on viscous heating.
Accordingly, $T_{\rm d}$ is computed as under the optically thick assumption \citep{rl86,nn94}:
\begin{equation}
\label{eq:T_d_thick}
T_{\rm d}^4  =  \frac{27 \tau}{128 \sigma_{\rm SB}}  \Sigma_{\rm g} \nu \Omega^2 
                     =  \frac{27 \kappa}{128 \sigma_{\rm SB}} \left( \frac{\dot{M}_{\rm d}}{3 \pi} \right)^2 \frac{\Omega^3}{ \alpha c_{\rm s}^2},
\end{equation}
where $\tau =\kappa \Sigma_{\rm g}$ is the optical depth, $\kappa$ is the opacity, and $\sigma_{\rm SB}$ is the Stefan-Boltzmann constant.
We also assume in equation (\ref{eq:T_d_thick}) that 
the vertical heat flux is constant and the temperature at the disk surface is negligible compared to that at the disk midplane \citep{hp11}.
Thus, $T_{\rm d}$ can be computed self-consistently for a given value of $\kappa$, using equations (\ref{eq:f_OR}) and (\ref{eq:T_d_thick}).

We follow \citet{bck97} to specify $\kappa$ that are functions of $\Sigma_{\rm g}$ and $T_{\rm d}$.
In the preliminary study, 
we considered all the opacity regimes summarized in Table 2 of \citet[from $n=1$ to $n=9$, where $n$ is the opacity regime index]{bck97} and 
found that the results of all the cases are bracketed by those of $n=7$ and $n=8$ within the water snow line.
The former case corresponds to the opacity originating from metal grains,
while the latter corresponds to the opacity regime where metal grains sublimate due to high temperatures.
Here, we focus only on these two opacity regimes, 
which are hereafter referred to as the metal grain and the sublimation cases, respectively.

In summary, there are three parameters in our model: $k_{\rm OR}$, $\dot{M}_{\rm d}$, and $M_{p}$.
We simply assume that $k_{\rm OR}=1$.
For $\dot{M}_{\rm d}$, we adopt $10^{-8} M_{\odot}$ yr$^{-1}$ based on disk observations \citep{wc11}.
We set $M_{p}$ to $10 M_{\oplus}$ because this value may represent the typical core mass (see Section \ref{sec:disc} for more discussion).
We note that variation of these parameters does not affect the profiles of the disk properties; variation only causes shifts in the absolute magnitude.

\subsection{Resulting disk properties}  \label{sec:mod_3}

\begin{figure*}
\begin{center}
\includegraphics[width=6cm]{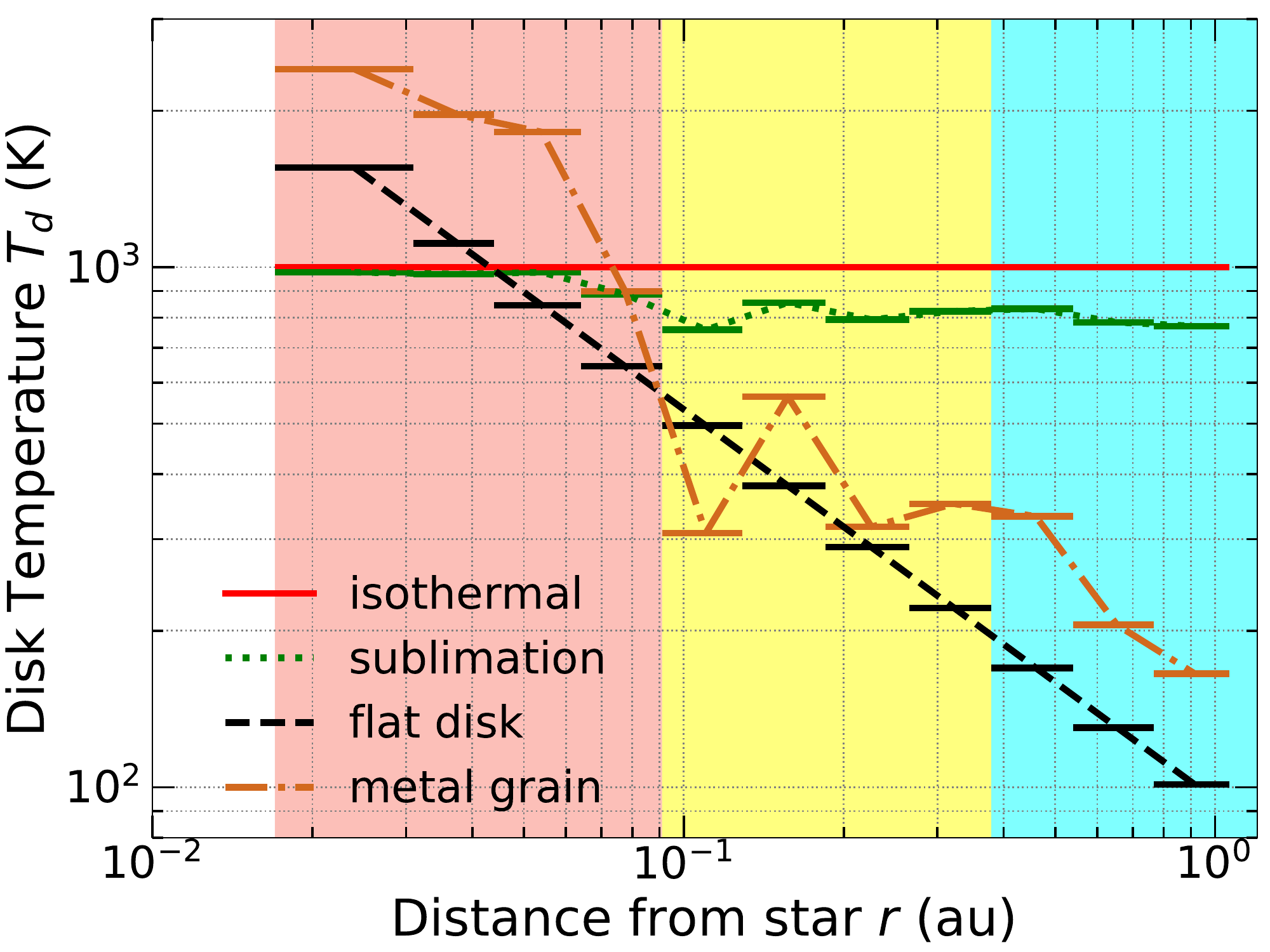}
\includegraphics[width=6cm]{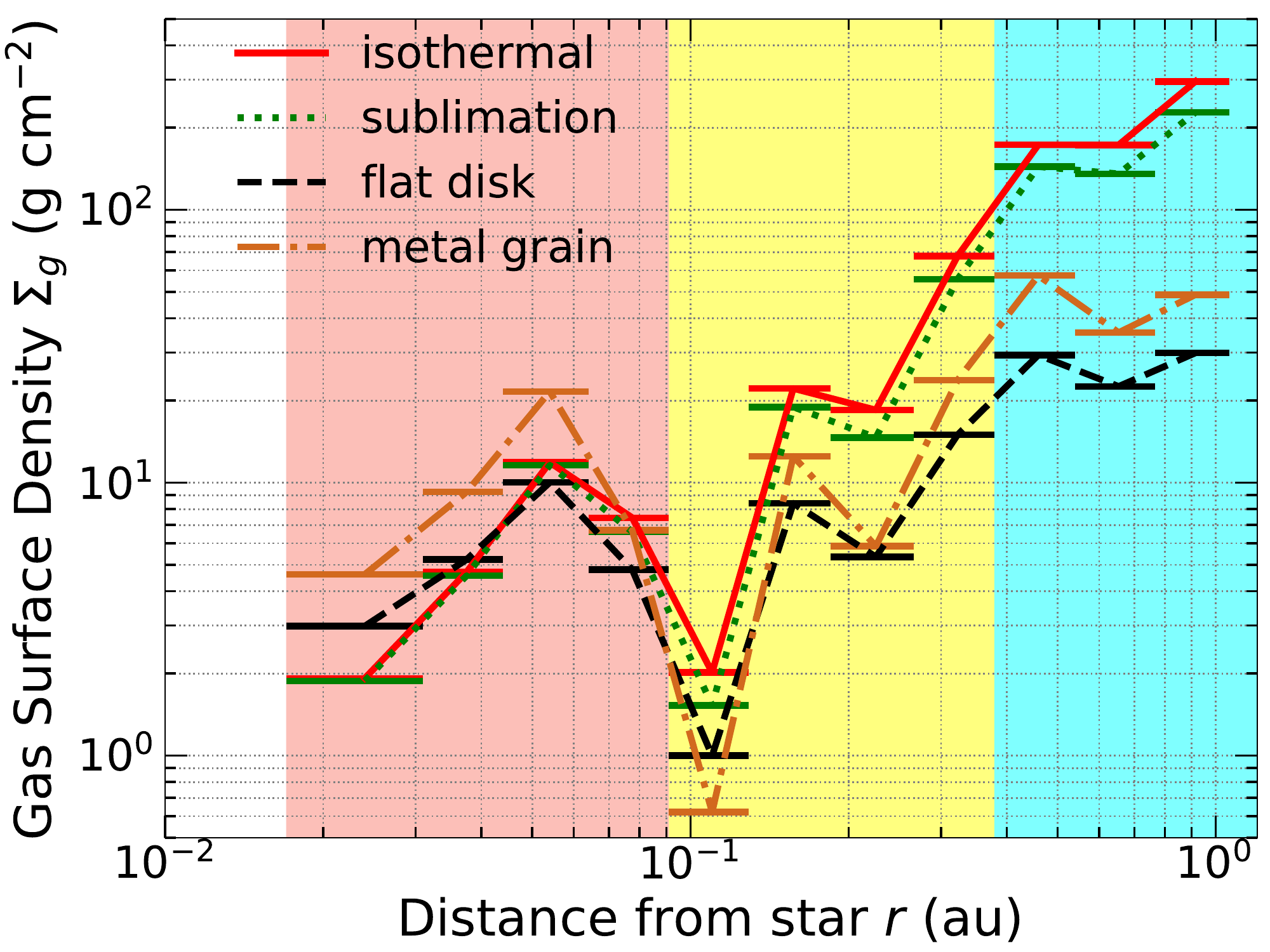}
\includegraphics[width=6cm]{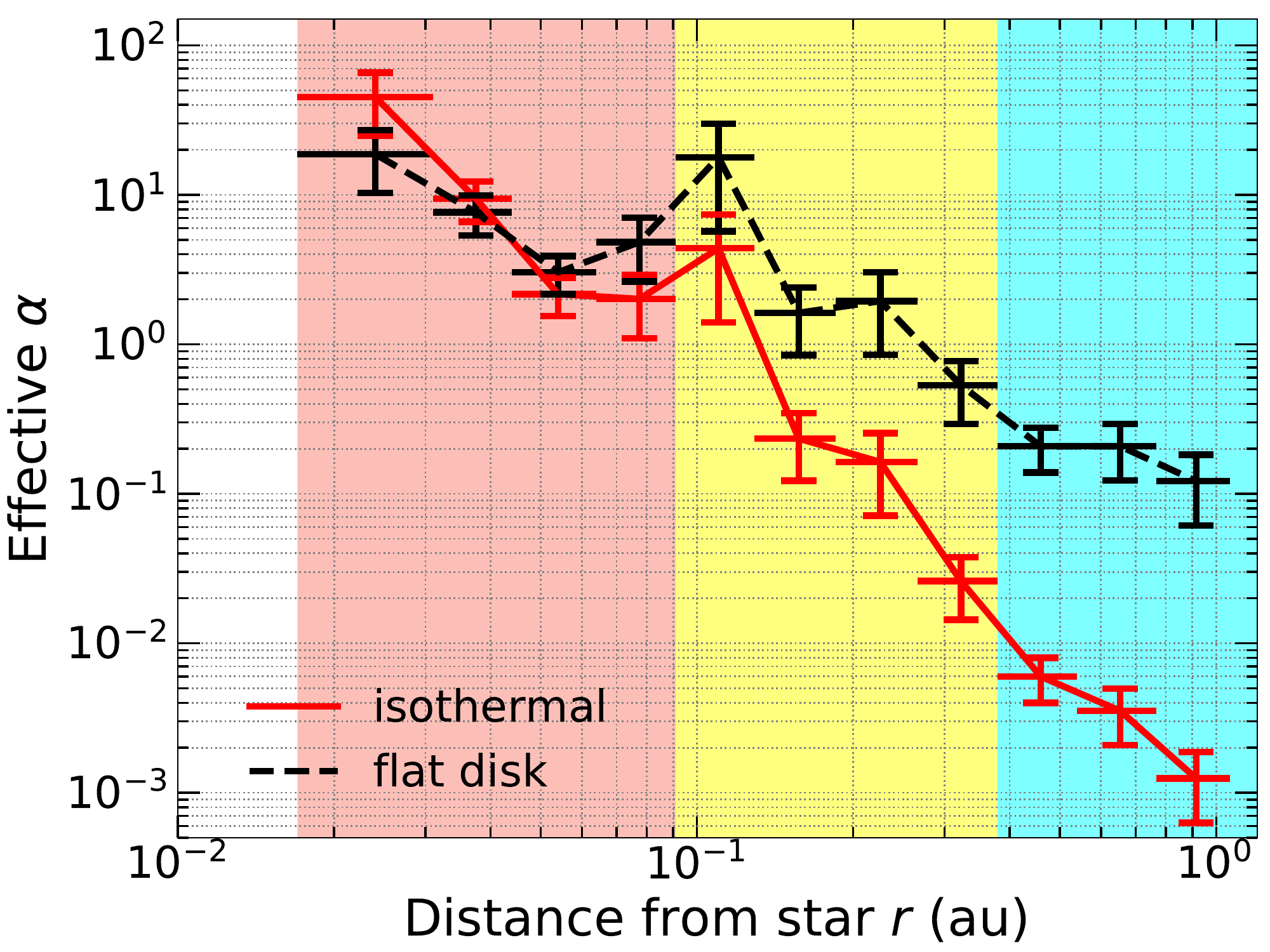}
\caption{Resulting structures of $T_{\rm d}$, $\Sigma_{\rm g}$, and $\alpha$ 
as a function of the distance from the central star in the left, central, and right panels, respectively.
The results for the isothermal case are denoted by the red solid line, those for the sublimation case by the green dotted line,
those for the flat disk case by the black dashed line, and those for the metal grain case by the orange dash-dotted line.
As in Figure \ref{fig1}, the horizontal bars are displayed while error bars are omitted for the purpose of clear presentation in the left and central panels.
Both the horizontal bars and error bars are displayed in the right panel.}
\label{fig2}
\end{center}
\end{figure*}

We present all the disk properties ($T_{\rm d}$, $\Sigma_{\rm g}$, and $\alpha$) that are computed self-consistently.
For comparison purposes, we also consider two given temperature profiles:
 the isothermal case ($T_{\rm d}=10^3$ K), 
and the flat disk case: $T_{\rm d} \simeq ( 2/(3 \pi) )^{1/4} ( R_*/r )^{3/4} T_*$,  
where $R_*=1.5R_{\odot}$ and $T_*=5780$ K are the radius and temperature of the central star, respectively \citep{cg97}.
We note that $T_{\rm d}$ is determined by stellar irradiation for the flat disk case.

Figure \ref{fig2} shows our results for $T_{\rm d}$, $\Sigma_{\rm g}$, and $\alpha$.
We find that the radial variation of $T_{\rm d}$ is small for the sublimation case.
Interestingly, this behavior is very similar to that seen in the isothermal case.
For the metal grain case, $T_{\rm d}$ becomes a roughly decreasing function of $r$.
This trend is broadly consistent with the results of the flat disk case.
We confirm similar features for $\Sigma_{\rm g}$.
We can therefore conclude that for the opacity regime currently considered,
the isothermal assumption works well for the sublimation case and
the flat disk model provides reasonable results for the metal grain case.
In the following, we consider only the isothermal and the flat disk cases.
Our results also show that $\Sigma_{\rm g}$ traces the shape of $f_{\rm OR}$ for all cases considered here (see Figure \ref{fig1}).
The in-situ gas accretion hypothesis can therefore reproduce the occurrence rate of observed close-in planets 
if the gas surface density of disks has a similar distribution.
We now discuss the $\alpha$ profile (see right panel).
We observe that the resulting $\alpha$ profile is characterized by $1/ \Sigma_{\rm g}$,
which is expected under the steady state disk model.
We note that these disk properties should be viewed as the mean values 
due to plausible differences in core mass and formation timing of close-in Jovian planets.

In the following, we use the above disk properties to compute the radial distribution of magnetic fields threading disks.

\subsection{Magnetic field profile}  \label{sec:mod_4}

Magnetic fields threading disks play a key role in disk evolution \citep[for a review]{nfg14}.
Recent studies make it evident that the angular momentum transport of disks and the resulting $\alpha$ are regulated by both nonideal MHD effects and vertical magnetic flux.
As shown in Figure \ref{fig2}, however, the inner region of disks becomes hot ($\ga 800$ K) enough for thermal and collisional ionization to be effective \citep{g96}.
This indicates that the ideal MHD limit would still work well for most parts of inner disks.
We therefore consider both ideal MHD and nonideal MHD cases here.

We use the results of detailed MHD simulations.
For the ideal MHD case, the angular momentum of the disk is transported radially by magnetorotational instability (MRI) and the resulting MHD turbulence \citep{bh98}.
We adopt the $\alpha$-expression given by  \citet{ssa16}:
\begin{equation}
\label{eq:alpha_S16}
\alpha_{\rm S16} = 11 \beta_z^{-0.53},
\end{equation}
where $\beta_z = (\rho_{\rm g} c_{\rm s}^2)/(B_z^2/8\pi)$ is the plasma beta, $\rho_{\rm g}= \Sigma_{\rm g}/ (\sqrt{2\pi}H_{\rm g})$ is the gas density at the disk midplane, 
and $B_z$ is the net vertical magnetic flux threading disks.

When nonideal MHD effects (Ohmic and ambipolar diffusion and Hall term) come into play, 
the picture of disk evolution changes considerably.
In fact, recent nonideal MHD simulations reveal that 
inclusion of Ohmic and ambipolar diffusion with large-scale magnetic fields can lead to the quench of MRI-induced turbulence 
and the launch of disk winds that remove the angular momentum of the disk in the vertical direction \citep{bs13,b13}.
When all three nonideal MHD terms are included,
the results become sensitive to the direction of magnetic fields (aligned or anti-aligned with the rotation axis of the disk)
while disks tend to be laminar and disk winds still play a considerable role in disk evolution \citep[][]{b14,lkf14}.
The magnetic field configuration relative to the rotation axis of the disk is unknown, 
and the results with Ohmic and ambipolar diffusion provide the intermediate behavior 
between the results of full nonideal MHD simulations with the aligned and anti-aligned cases \citep{b13,b14}.
Therefore,  here we use the results of nonideal MHD simulations with Ohmic and ambipolar diffusion.
Assuming that angular momentum is removed largely by disk winds vertically in almost laminar disks,
equation (\ref{eq:mdot}) can be modified as \citep{hof17}
\begin{equation}
\label{eq:mdot_wind}
\dot{M}_{\rm d} \simeq  4 \sqrt{2 \pi} r \Sigma_{\rm g} c_{\rm s} W_{\rm z\phi},
,\end{equation}
where $W_{\rm z\phi}$ is the normalized accretion stress originating from disk winds.
We adopt the functional form of $W_{\rm z\phi}$ obtained by \citet{b13}:
\begin{equation}
\label{eq:w_zphi}
W_{\rm z\phi}^{\rm B13} = 0.23 \left( \frac{r}{1 \mbox{ au}} \right)^{0.46} \beta_z^{-0.66}.
\end{equation}
Consequently, the effective $\alpha$ can be written as
\begin{equation}
\label{eq:alpha_b13}
\alpha_{\rm B13} \equiv \frac{4r}{3 \sqrt{\pi}H_{\rm g}} W_{\rm z\phi}^{\rm B13}.
\end{equation}

\begin{figure*}
\begin{minipage}{17cm}
\begin{center}
\includegraphics[width=8cm]{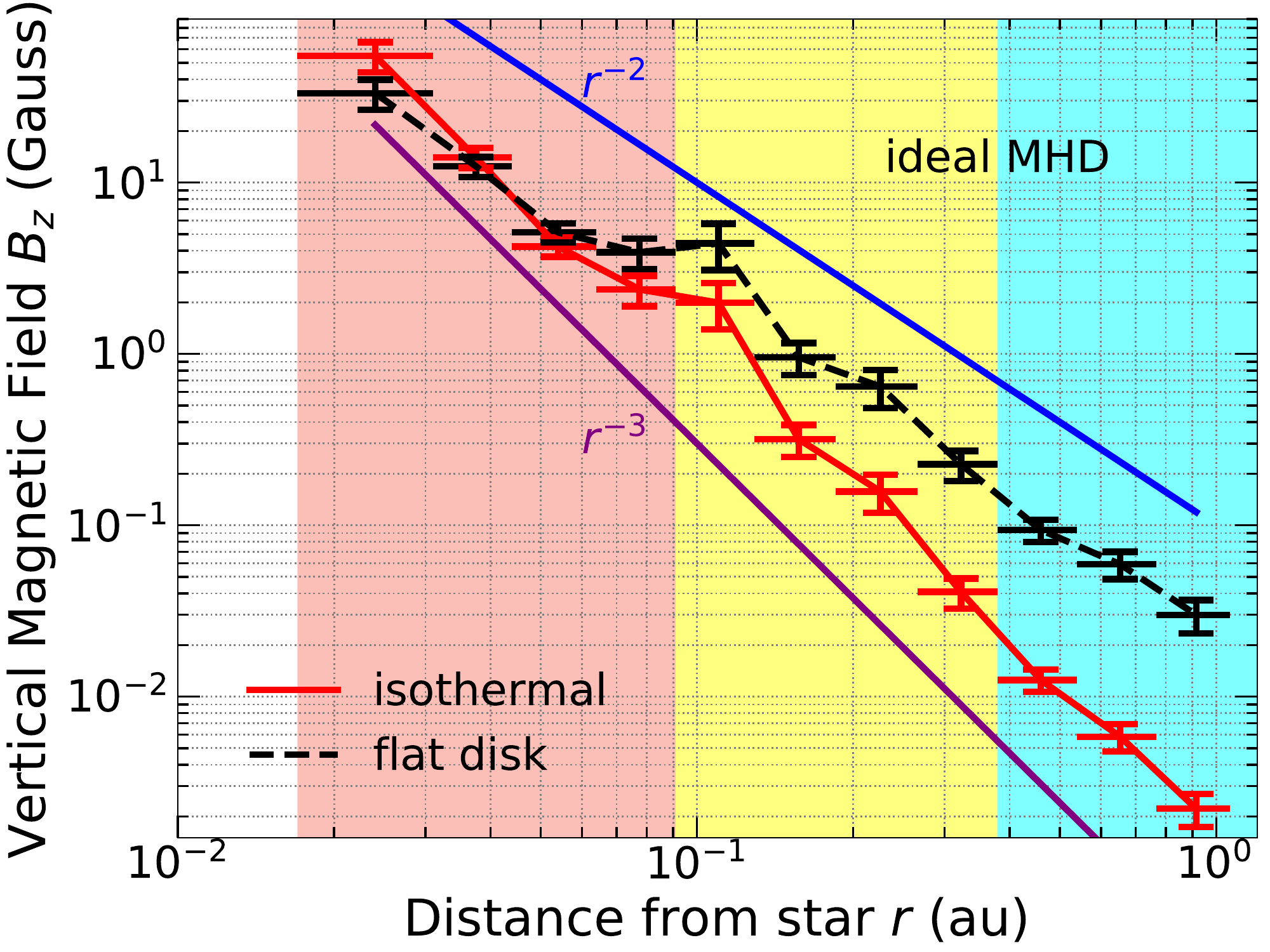}
\includegraphics[width=8cm]{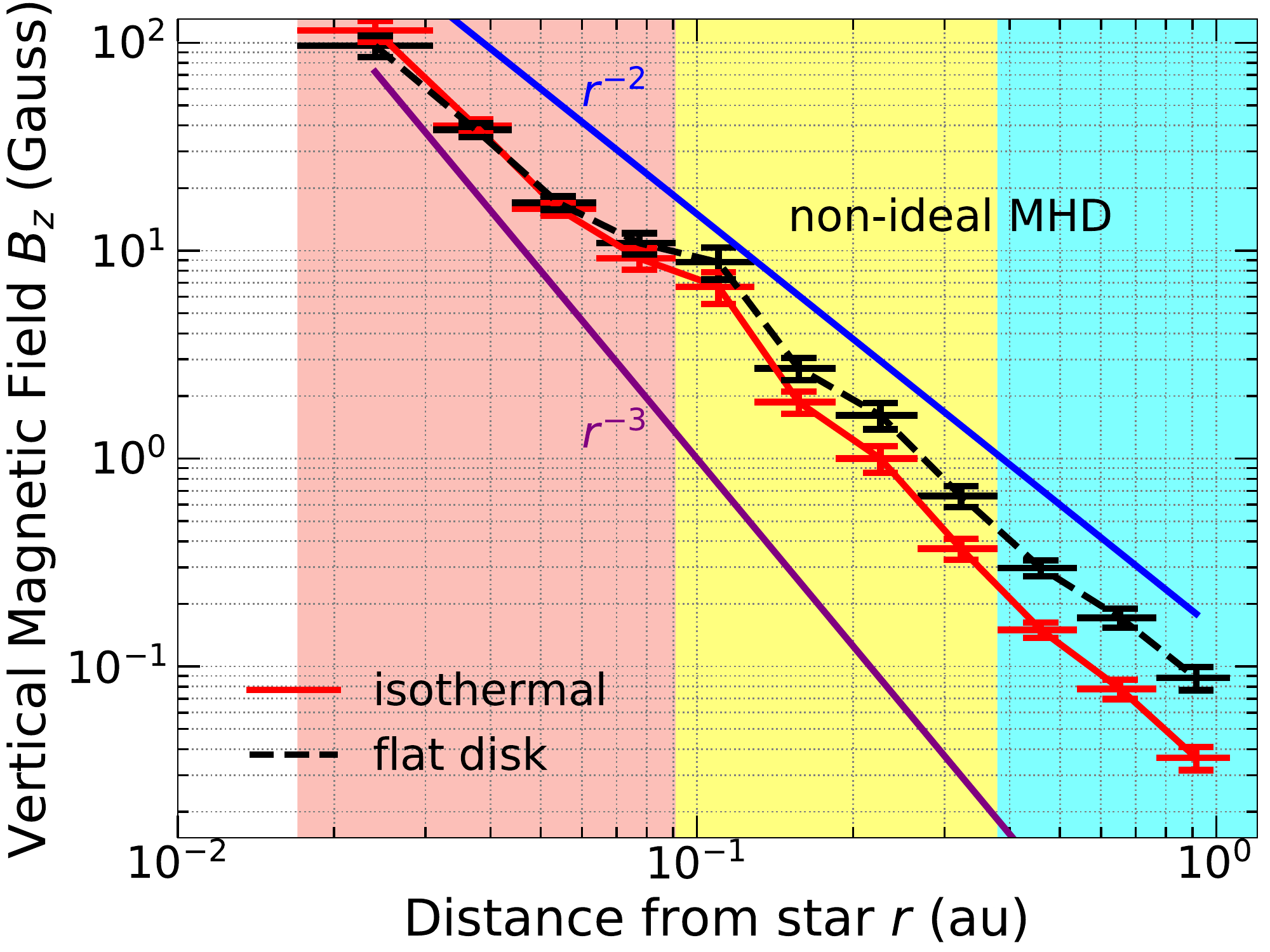}
\caption{Resulting magnetic field profiles as a function of $r,$ as in the right panel of Figure \ref{fig2}.
The results for the ideal MHD case are shown in the left panel, and those for the nonideal MHD case are shown in the right panel.
For reference, the profiles of $r^{-2}$ and $r^{-3}$ are denoted by the blue and the purple solid lines, respectively.}
\label{fig4}
\end{center}
\end{minipage}
\end{figure*}

We now compute the magnetic field profiles ($B_z$) that are reconstructed from the $\alpha$ profiles (Figure \ref{fig2}), 
using equations (\ref{eq:alpha_S16}) and (\ref{eq:alpha_b13}).
Figure \ref{fig4} shows our results.
We find that there are plateaus in the resulting magnetic field profiles around $r\simeq 0.1$ au for all cases.
Importantly, these locations correspond to the transition between hot Jupiters and giant planets in the period valley.
Our results also show that except for the isothermal case in the ideal MHD limit,
$B_z$ switches from $r^{-3}$ to $r^{-2}$ as $r$ increases.
It is clear that the switch in profiles originates from $f_{\rm OR}$.

Therefore, the occurrence rate distribution can be converted to disk properties ($\Sigma_{\rm g}$, $\alpha$, and $B_z$)
under our hypothesis.

\section{Discussion} \label{sec:disc}

We first point out that the resulting profiles of $\Sigma_{\rm g}$ and $\alpha$ are quite different from those generated by the standard model:
our $\Sigma_{\rm g}$ is a roughly increasing function of $r$ with some structures (Figure \ref{fig2}),
while the minimum-mass solar nebula (MMSN) model goes to $r^{-3/2}$ \citep{h81}.
The gas distribution in the inner region of disks is poorly constrained by the MMSN model due to the absence of gas giants there.
It is worth mentioning that the importance of increasing surface density profiles with $r$ has already been recognized in the literature. 
\citet{omg15} suggest that with such a profile, the population of observed close-in super-Earths can be better reproduced.
We note that they assume that the origin of the profile is mass loss originating from disk winds without considering magnetic field profiles.
Our results would be reasonable within $\la 1$ au because the opacity regimes currently considered are effective there;
other opacity regimes become important beyond this region, which changes the gas surface density and disk temperature profiles, and the importance of nonideal MHD effects.
The recent analysis of radial velocity observations suggests that the occurrence rate of giant planets may have a peak at $r=2-3$ au \citep{fmp19}.
This implies that other formation mechanisms would operate there as well.

For $\alpha$, our profile broadly predicts a decreasing function of $r$ with some features around $r \simeq 0.1$ au (Figure \ref{fig2}).
In particular, one would notice that $\alpha \ga 1$ in the vicinity of the central star.
We remind the readers that the absolute values of disk quantities are scaled by model parameters ($k_{\rm OR}$, $M_p$, and $\dot{M}_{\rm d}$)
and that $k_{\rm OR}$ is totally unknown.
For reference, we have computed the plasma $\beta_z$  value with equations (\ref{eq:alpha_S16}) and (\ref{eq:w_zphi})
and confirmed that under the current setup,
the values of $\alpha$ at $r \ga 0.04$ au and at $r \ga 0.4$ au are achieved by recent ideal and nonideal MHD simulations, respectively \citep{ssa16,b13}.

The most intriguing quantity would be magnetic field profiles.
While it is very difficult to infer the magnetic field profile observationally,
a number of theoretical studies are available.
For instance, it is well recognized that disk inner edges may be threaded by stellar magnetic fields
and the resulting magnetospheres would be important for disk evolution in the vicinity of classical T Tauri stars \citep{gl79,k91}.
If the dipole approximation is reasonable, 
the stellar vertical magnetic flux ($B_{\rm s}$) threading disks can be written as \citep[e.g.,][]{a16}
\begin{equation}
\label{eq:Bs}
B_{\rm s}  \simeq  10^3 \left( \frac{r}{1.5 R_{\odot}} \right)^{-3} \mbox{ G}  \simeq   41 \left( \frac{r}{2 \times 10^{-2} \mbox{ au}} \right)^{-3} \mbox{ G}.
\end{equation}
We note that the averaged stellar surface magnetic field strength is adopted in equation (\ref{eq:Bs}),
which is on the order of $10^3$ G for classical T Tauri stars \citep[e.g.,][]{j07}.
Such strength is probably needed for truncating the disk inner edge and making magnetospheric accretion effective \citep{k91}.
However, spectropolarimetry observations suggest that quadrupolar and higher-order moments provide the dominant contribution to the strength \citep[e.g.,][]{dmp08},
implying that the dipole field may be much weaker \citep[for a review]{hhc16}.
In fact, \citet{rt05a} adopt a weak stellar field and compute the disk properties in detail \citep[also see][]{rt05b}.  
Therefore, equation (\ref{eq:Bs}) should be viewed as an upper limit for stellar dipole fields.
In addition,
protoplanetary disks would be threaded by large-scale magnetic fields ($B_{\rm d}$) that originate from interstellar fields
that are the consequence of disk formation and evolution inherited from star formation \citep{lpp94,otm14}.
For this case, $B_{\rm d}$ is determined by two competing processes: advection and diffusion of magnetic flux in the radial direction.
Adopting the steady state solution,
$B_{\rm d}$ can be  given as \citep{otm14}
\begin{equation}
\label{eq:B_d}
B_{\rm d} \simeq 0.1 \left( \frac{r}{1 \mbox{ au}} \right)^{-2} \mbox{ G}.
\end{equation}
This should be viewed as an upper limit for the large-scale magnetic fields of disks because it is derived from the assumption of highly conducting accretion disks.
Interestingly, our results of $B_z$ are comparable with these two upper limits (Figure \ref{fig4}).
This work therefore implies that stellar fields might affect the formation of hot Jupiters
while large-scale fields might be important for giant planet formation beyond $r \simeq 0.1$ au.
Furthermore, switching these two fields might lead to the period valley in $f_{\rm OR}$.

How do our disk properties affect planetary migration?
As shown by previous work, type I migration is halted at the disk inner edge \citep{mmcf06,omg15}.
For type II migration, its speed can be reduced significantly 
if the planet mass at $r=r_p$ exceeds the local disk mass \citep[$\Delta M_{\rm d} \simeq 2 \pi \Sigma_{\rm g}r_p^2$, e.g.,][]{hi13}.
At $r_p \simeq 1$ au, $\Sigma_{\rm g} \sim 10^2$ g cm$^{-2}$ and $\Delta M_{\rm d} \sim 24M_{\oplus}$ for our current setup.
Thus, both type I and II migration would be negligible in our model, which justifies the assumption of the in-situ scenario.

We now list the potential issues contained in this work.
The most critical one is that we have adopted $M_p = 10 M_{\oplus}$ in equation (\ref{eq:f_OR}).
The presence of such massive cores in the inner region of disks is fundamental for the in-situ gas accretion scenario.
Nonetheless, equation (\ref{eq:f_OR}) implicitly assumes that the formation frequency of such cores may be uniform there.
Since {\it Kepler} observations show that the populations of both super-Earths and sub-Neptunes increase monotonically from 1 day to 10 days and
become relatively flat beyond 10 days \citep[e.g.,][]{pmw18},
our assumption would be justified only for longer periods.
In other words, low $f_{\rm OR}$ within 10 days can be caused by either low $\Sigma_{\rm g}$ or low abundance of planetary cores.
We also note that the {\it Kepler} data lead to a similar occurrence rate distribution even with different analyses \citep{smt16,pmw18}.
On the contrary, the analysis of only radial velocity data tends to gain a higher value of the occurrence rate for close-in Jovian planets \citep{mml11}.
More observations and/or systematic analyses are needed to examine the difference.

We finally discuss how close-in giant planets form.
Our results clearly show that the in-situ gas accretion can work for planets located beyond 0.1 au.
For planets within 0.1 au, it may not be impossible to form some fraction of hot Jupiters via the in-situ gas accretion scenario as suggested by \citet{bhl00} and this work.
However, the scenario would entail certain problems, such as the need to populate massive cores and/or unreasonably high values of $\alpha$ there.
Instead, other scenarios may provide reasonable explanations for such planets.
For instance, the peak value of $f_{\rm OR}$ at $\sim 4$ days can be understood by the combination of high-eccentricity migration and tidal circularization \citep{wl11},
and other orbital features such as obliquity can be explained by the Lidov-Kozai effect \citep[e.g.,][]{n16}.

In the near future, better modeling and observations of inner disk regions will become available to directly test our prediction and 
eventually to reveal the appearance of natal protoplanetary disks and how planet formation takes place in the vicinity of central stars.

\begin{acknowledgements}
The authors thank an anonymous referee for useful comments on our manuscript.
Y.H. thanks Konstantin Batygin and Trevor David for stimulating discussions.
This research was carried out at JPL/Caltech under a contract with NASA.
Y.H. is supported by JPL/Caltech.
\end{acknowledgements}




\begin{thebibliography}{57}
\expandafter\ifx\csname natexlab\endcsname\relax\def\natexlab#1{#1}\fi

\bibitem[{Armitage(2011)}]{a11}
Armitage, P.~J. 2011, \araa, 49, 195

\bibitem[{{Armitage}(2016)}]{a16}
{Armitage}, P.~J. 2016, \apjl, 833, L15

\bibitem[{{Bai}(2013)}]{b13}
{Bai}, X.-N. 2013, \apj, 772, 96

\bibitem[{{Bai}(2014)}]{b14}
{Bai}, X.-N. 2014, \apj, 791, 137

\bibitem[{{Bai} \& {Stone}(2013)}]{bs13}
{Bai}, X.-N. \& {Stone}, J.~M. 2013, \apj, 769, 76

\bibitem[{{Balbus} \& {Hawley}(1998)}]{bh98}
{Balbus}, S.~A. \& {Hawley}, J.~F. 1998, Reviews of Modern Physics, 70, 1

\bibitem[{{Batygin} {et~al.}(2016){Batygin}, {Bodenheimer}, \&
  {Laughlin}}]{bbl16}
{Batygin}, K., {Bodenheimer}, P.~H., \& {Laughlin}, G.~P. 2016, \apj, 829, 114

\bibitem[{{Bell} {et~al.}(1997){Bell}, {Cassen}, {Klahr}, \& {Henning}}]{bck97}
{Bell}, K.~R., {Cassen}, P.~M., {Klahr}, H.~H., \& {Henning}, T. 1997, \apj,
  486, 372

\bibitem[{{Bodenheimer} {et~al.}(2000){Bodenheimer}, {Hubickyj}, \&
  {Lissauer}}]{bhl00}
{Bodenheimer}, P., {Hubickyj}, O., \& {Lissauer}, J.~J. 2000, Icarus, 143, 2

\bibitem[{{Boley} {et~al.}(2016){Boley}, {Granados Contreras}, \&
  {Gladman}}]{bgg16}
{Boley}, A.~C., {Granados Contreras}, A.~P., \& {Gladman}, B. 2016, \apjl, 817,
  L17

\bibitem[{{Chiang} \& {Laughlin}(2013)}]{cl13}
{Chiang}, E. \& {Laughlin}, G. 2013, \mnras, 431, 3444

\bibitem[{Chiang \& Goldreich(1997)}]{cg97}
Chiang, E.~I. \& Goldreich, P. 1997, \apj, 490, 368

\bibitem[{D{'}Alessio {et~al.}(1998)D{'}Alessio, Cant{\'o}, Calvet, \&
  Lizano}]{dccl98}
D{'}Alessio, P., Cant{\'o}, J., Calvet, N., \& Lizano, S. 1998, \apj, 500, 411

\bibitem[{{Dawson} \& {Johnson}(2018)}]{dj18}
{Dawson}, R.~I. \& {Johnson}, J.~A. 2018, \araa, 56, 175

\bibitem[{{Donati} {et~al.}(2008){Donati}, {Morin}, {Petit}, {Delfosse},
  {Forveille}, {Auri{\`e}re}, {Cabanac}, {Dintrans}, {Fares}, {Gastine},
  {Jardine}, {Ligni{\`e}res}, {Paletou}, {Ramirez Velez}, \&
  {Th{\'e}ado}}]{dmp08}
{Donati}, J.~F., {Morin}, J., {Petit}, P., {et~al.} 2008, \mnras, 390, 545

\bibitem[{{Fernandes} {et~al.}(2019){Fernandes}, {Mulders}, {Pascucci},
  {Mordasini}, \& {Emsenhuber}}]{fmp19}
{Fernandes}, R.~B., {Mulders}, G.~D., {Pascucci}, I., {Mordasini}, C., \&
  {Emsenhuber}, A. 2019, \apj, 874, 81

\bibitem[{{Frank} {et~al.}(2002){Frank}, {King}, \& {Raine}}]{fkr02}
{Frank}, J., {King}, A., \& {Raine}, D.~J. 2002, {Accretion Power in
  Astrophysics: Third Edition}

\bibitem[{Gammie(1996)}]{g96}
Gammie, C.~F. 1996, \apj, 457, 355

\bibitem[{{Ghosh} \& {Lamb}(1979)}]{gl79}
{Ghosh}, P. \& {Lamb}, F.~K. 1979, \apj, 232, 259

\bibitem[{{Hartmann} {et~al.}(2016){Hartmann}, {Herczeg}, \& {Calvet}}]{hhc16}
{Hartmann}, L., {Herczeg}, G., \& {Calvet}, N. 2016, \araa, 54, 135

\bibitem[{{Hasegawa} {et~al.}(2018){Hasegawa}, {Bryden}, {Ikoma}, {Vasisht}, \&
  {Swain}}]{hbi18}
{Hasegawa}, Y., {Bryden}, G., {Ikoma}, M., {Vasisht}, G., \& {Swain}, M. 2018,
  \apj, 865, 32

\bibitem[{{Hasegawa} {et~al.}(2019){Hasegawa}, {Hansen}, \& {Vasisht}}]{hhv19}
{Hasegawa}, Y., {Hansen}, B. M.~S., \& {Vasisht}, G. 2019, \apjl, 876, L32

\bibitem[{{Hasegawa} \& {Ida}(2013)}]{hi13}
{Hasegawa}, Y. \& {Ida}, S. 2013, \apj, 774, 146

\bibitem[{{Hasegawa} {et~al.}(2017){Hasegawa}, {Okuzumi}, {Flock}, \&
  {Turner}}]{hof17}
{Hasegawa}, Y., {Okuzumi}, S., {Flock}, M., \& {Turner}, N.~J. 2017, \apj, 845,
  31

\bibitem[{{Hasegawa} \& {Pudritz}(2011)}]{hp11}
{Hasegawa}, Y. \& {Pudritz}, R.~E. 2011, \mnras, 417, 1236

\bibitem[{{Hayashi}(1981)}]{h81}
{Hayashi}, C. 1981, Progress of Theoretical Physics Supplement, 70, 35

\bibitem[{{Johns-Krull}(2007)}]{j07}
{Johns-Krull}, C.~M. 2007, \apj, 664, 975

\bibitem[{{Kley} \& {Nelson}(2012)}]{kn12}
{Kley}, W. \& {Nelson}, R.~P. 2012, \araa, 50, 211

\bibitem[{{Koenigl}(1991)}]{k91}
{Koenigl}, A. 1991, \apjl, 370, L39

\bibitem[{{Lesur} {et~al.}(2014){Lesur}, {Kunz}, \& {Fromang}}]{lkf14}
{Lesur}, G., {Kunz}, M.~W., \& {Fromang}, S. 2014, \aap, 566, A56

\bibitem[{Lin {et~al.}(1996)Lin, Bodenheimer, \& Richardson}]{lbr96}
Lin, D. N.~C., Bodenheimer, P., \& Richardson, D.~C. 1996, \nat, 380, 606

\bibitem[{{Lubow} {et~al.}(1994){Lubow}, {Papaloizou}, \& {Pringle}}]{lpp94}
{Lubow}, S.~H., {Papaloizou}, J.~C.~B., \& {Pringle}, J.~E. 1994, \mnras, 267,
  235

\bibitem[{Masset {et~al.}(2006)Masset, Morbidelli, Crida, \& Ferreira}]{mmcf06}
Masset, F.~S., Morbidelli, A., Crida, A., \& Ferreira, J. 2006, \apj, 642, 478

\bibitem[{{Matsumura} {et~al.}(2010){Matsumura}, {Peale}, \& {Rasio}}]{mpr10}
{Matsumura}, S., {Peale}, S.~J., \& {Rasio}, F.~A. 2010, \apj, 725, 1995

\bibitem[{Mayor \& Queloz(1995)}]{mq95}
Mayor, M. \& Queloz, D. 1995, \nat, 378, 355

\bibitem[{Mayor {et~al.}(2011)}]{mml11}
Mayor, M. {et~al.} 2011, preprint (astro-ph/arXiv:1109.2497v1)

\bibitem[{{Morbidelli} {et~al.}(2014){Morbidelli}, {Szul{\'a}gyi}, {Crida},
  {Lega}, {Bitsch}, {Tanigawa}, \& {Kanagawa}}]{msc14}
{Morbidelli}, A., {Szul{\'a}gyi}, J., {Crida}, A., {et~al.} 2014, \icarus, 232,
  266

\bibitem[{{Mordasini} {et~al.}(2012){Mordasini}, {Alibert}, {Klahr}, \&
  {Henning}}]{mak12}
{Mordasini}, C., {Alibert}, Y., {Klahr}, H., \& {Henning}, T. 2012, \aap, 547,
  A111

\bibitem[{Nakamoto \& Nakagawa(1994)}]{nn94}
Nakamoto, T. \& Nakagawa, Y. 1994, \apj, 421, 640

\bibitem[{{Naoz}(2016)}]{n16}
{Naoz}, S. 2016, \araa, 54, 441

\bibitem[{{Ogihara} {et~al.}(2015){Ogihara}, {Morbidelli}, \&
  {Guillot}}]{omg15}
{Ogihara}, M., {Morbidelli}, A., \& {Guillot}, T. 2015, \aap, 584, L1

\bibitem[{{Okuzumi} {et~al.}(2014){Okuzumi}, {Takeuchi}, \& {Muto}}]{otm14}
{Okuzumi}, S., {Takeuchi}, T., \& {Muto}, T. 2014, \apj, 785, 127

\bibitem[{{Petigura} {et~al.}(2018){Petigura}, {Marcy}, {Winn}, {Weiss},
  {Fulton}, {Howard}, {Sinukoff}, {Isaacson}, {Morton}, \& {Johnson}}]{pmw18}
{Petigura}, E.~A., {Marcy}, G.~W., {Winn}, J.~N., {et~al.} 2018, \aj, 155, 89

\bibitem[{Pollack {et~al.}(1996)Pollack, Hubickyj, Bodenheimer, Lissauer,
  Podolak, \& Greenzweig}]{p96}
Pollack, J.~B., Hubickyj, O., Bodenheimer, P., {et~al.} 1996, Icarus, 124, 62

\bibitem[{Rasio \& Ford(1996)}]{rf96}
Rasio, F.~A. \& Ford, E.~B. 1996, Science, 274, 954

\bibitem[{Ruden \& Lin(1986)}]{rl86}
Ruden, S.~P. \& Lin, D. N.~C. 1986, \apj, 308, 883

\bibitem[{{Russo} \& {Thompson}(2015{\natexlab{a}})}]{rt05b}
{Russo}, M. \& {Thompson}, C. 2015{\natexlab{a}}, \apj, 815, 38

\bibitem[{{Russo} \& {Thompson}(2015{\natexlab{b}})}]{rt05a}
{Russo}, M. \& {Thompson}, C. 2015{\natexlab{b}}, \apj, 813, 81

\bibitem[{{Salvesen} {et~al.}(2016){Salvesen}, {Simon}, {Armitage}, \&
  {Begelman}}]{ssa16}
{Salvesen}, G., {Simon}, J.~B., {Armitage}, P.~J., \& {Begelman}, M.~C. 2016,
  \mnras, 457, 857

\bibitem[{{Santerne} {et~al.}(2016){Santerne}, {Moutou}, {Tsantaki}, {Bouchy},
  {H{\'e}brard}, {Adibekyan}, {Almenara}, {Amard}, {Barros}, {Boisse},
  {Bonomo}, {Bruno}, {Courcol}, {Deleuil}, {Demangeon}, {D{\'{\i}}az},
  {Guillot}, {Havel}, {Montagnier}, {Rajpurohit}, {Rey}, \& {Santos}}]{smt16}
{Santerne}, A., {Moutou}, C., {Tsantaki}, M., {et~al.} 2016, \aap, 587, A64

\bibitem[{{Shakura} \& {Sunyaev}(1973)}]{ss73}
{Shakura}, N.~I. \& {Sunyaev}, R.~A. 1973, \aap, 24, 337

\bibitem[{{Tanigawa} \& {Ikoma}(2007)}]{ti07}
{Tanigawa}, T. \& {Ikoma}, M. 2007, \apj, 667, 557

\bibitem[{{Tanigawa} \& {Watanabe}(2002)}]{tw02}
{Tanigawa}, T. \& {Watanabe}, S.-i. 2002, \apj, 580, 506

\bibitem[{{Turner} {et~al.}(2014){Turner}, {Fromang}, {Gammie}, {Klahr},
  {Lesur}, {Wardle}, \& {Bai}}]{nfg14}
{Turner}, N.~J., {Fromang}, S., {Gammie}, C., {et~al.} 2014, Protostars and
  Planets VI, 411

\bibitem[{Williams \& Cieza(2011)}]{wc11}
Williams, J.~P. \& Cieza, L.~A. 2011, \araa, 49, 67

\bibitem[{{Winn} \& {Fabrycky}(2015)}]{wf15}
{Winn}, J.~N. \& {Fabrycky}, D.~C. 2015, \araa, 53, 409

\bibitem[{{Wu} \& {Lithwick}(2011)}]{wl11}
{Wu}, Y. \& {Lithwick}, Y. 2011, \apj, 735, 109

\end{thebibliography}

\end{document}